\documentclass[aps,pre,twocolumn,superscriptaddress]{revtex4}

\usepackage{amsfonts,amssymb,amsmath,latexsym,epsfig,mathrsfs}
\usepackage{graphicx}
\usepackage{color}
\usepackage{subfigure}
\usepackage{url} 
\usepackage{hyperref}

\begin{document}              

\title{Diffusion in randomly perturbed dissipative dynamics \\ EPL 108 40002 (2014) - doi:10.1209/0295-5075/108/40002}

\author{Christian S. Rodrigues}~\email{christian.rodrigues@mis.mpg.de}
\affiliation{Max Planck Institute for Mathematics in the Sciences, Inselstr., 22, 04103 Leipzig, Germany}

\author{Aleksei V. Chechkin}
\affiliation{Institute for Theoretical Physics, NSC KIPT, ul.\ Akademicheskaya 1, UA-61108 Kharkov, Ukraine}

\author{Alessandro P. S. de Moura}
\affiliation{Department of Physics and Institute for Complex Systems and Mathematical Biology, King's College, University of Aberdeen - Aberdeen AB24 3UE, UK}

\author{Celso Grebogi} 
\affiliation{Department of Physics and Institute for Complex Systems and Mathematical Biology, King's College, University of Aberdeen - Aberdeen AB24 3UE, UK}

\author{Rainer Klages}
\affiliation{Queen Mary University of London, School of Mathematical Sciences, Mile End Road - London E1 4NS, UK}

\date{\today}

\begin{abstract}

\noindent
Dynamical systems having many coexisting attractors present
interesting properties from both fundamental theoretical and modelling
points of view. When such dynamics is under bounded random
perturbations, the basins of attraction are no longer invariant and
there is the possibility of transport among them. Here we introduce a
basic theoretical setting which enables us to study this hopping
process from the perspective of anomalous transport using the concept
of a random dynamical system with holes. We apply it to a simple model
by investigating the role of hyperbolicity for the transport among
basins. We show numerically that our system exhibits non-Gaussian
position distributions, power-law escape times, and subdiffusion. Our
simulation results are reproduced consistently from stochastic
Continuous Time Random Walk theory.

\end{abstract}

\pacs{05.40.-a}
\pacs{05.60.-k}
\pacs{05.45.-a}

\maketitle

\section{\label{sec.Int} Introduction}

Understanding the dynamics of systems exhibiting coexisting attractors
is fundamental for modelling processes having many possible asymptotic
states. Although not restricted to, this multi-stable dynamics is
particularly important in systems experiencing very weak
dissipation~\cite{FGH96, RMG09}. In contrast to strongly dissipative
ones, these are typically not dominated by one or few
attractors. There are many areas from which we could pick up such
examples. For instance, if one considers finite-size particle in
advection dynamics, the low dissipative interaction between the
advected particles and the fluid can be characterised by the presence
of multiple attractors trapping advected particles even in open
flows~\cite{ViM07}. Another example is found in the dynamics of space
dust and its role in the formation of planetesimals~\cite{BST}, among
others. Even when most of the attracting sets are periodic, a chaotic
component may be present in the form of a fractal boundary separating
the basins of attraction~\cite{FGH96, RMG09}. If the dynamics is fully
deterministic, the attractors are invariant structures. Hence, once a
particle or trajectory is trapped in one of the basins of attraction,
it remains there indefinitely. However, since most natural processes
are not realistically isolated from external random perturbations, it
is natural to study their impact.

The presence of random noise dramatically changes the dynamics. In
contrast to deterministic systems, for randomly perturbed dynamics the
invariance of attractors may not be true. If the considered
perturbation is set to be unbounded Gaussian noise, the whole phase
space may be the support of a unique invariant
measure~\cite{Arn98}. When bounded perturbations are used, on the
other hand, there might be many coexisting invariant measures. In
particular, depending on the amplitude of the noise orbits can escape
from the attracting domains~\cite{RGM10} creating the possibility of
transport across their basins. This sort of hopping process has been
reported before~\cite{BBS02, FGH96, LaT11, ZmH07}, yet there is a lack
of understanding of its statistical properties, in particular from the
anomalous transport perspective \cite{KRS08,MKl00}.

In this paper we analyse the statistical properties of systems lying
on the border between dissipative and conservative dynamics which
evolve under random perturbations and their similarities to
Hamiltonian dynamics. We start by introducing what we call
\textit{effective attractors}.  Below a certain level of dissipation
the dynamics naturally gives rise to these attracting sets, which
defined under finite resolution are indistinguishable from topological
attractors. We then extend the description of escape in terms of a
closed systems with a hole~\cite{RGM10} to the case of coexisting
attractors and establish the conditions allowing a hopping dynamics
among them. We show that it is possible to characterise the hopping
process by a distribution of first recurrence times to an
appropriately chosen non-zero measure set. We find that such a
recurrence (or escape time) distribution approaches the one expected
for non-hyperbolic dynamics as the dissipation is decreased and the
dynamics approaches the non-hyperbolic limit. This effect is similar
to stickiness in Hamiltonian non-hyperbolic dynamics
\cite{Ott02,KRS08}. We verify our arguments by computer simulations
for the single rotor or dissipative standard map
\cite{Zas78,FGH96}. The results match well to analytical predictions
from stochastic Continuous Time Random Walk theory
\cite{GeTo84,MKl00,Kor07}. Our discussion is based on general
arguments and not restricted to this particular model.

\section{\label{sec.fram}Dynamics and effective attractors}

We are interested in adding bounded random noise to our deterministic
dynamics.  More precisely, suppose our deterministic dynamics is given
by the iteration of a smooth function $f : M \to M$ with
differentiable inverse in our phase space $M$, for
example,\footnote{More generally, $M$ is a Riemannian manifold.} $M
\subset \mathbb{R}^{n}$. An orbit $(x_{n})_{n\geq1}$ is the sequence
generated by the dynamical system $x_{n+1} = f(x_{n})$ from a given
initial condition $x_{0} \in M$. We will be concerned with subsets of
$M$ to which most orbits in their neighbourhood converge for
sufficiently long but \textit{finite time}, what we shall call
\textit{effective attractors} or attracting sets. In other words,
those are $f$-invariant subsets of M contained in basins of
attraction, which are open sets of initial conditions with positive
Lebesgue (volume) measure converging to the attracting sets. Note that
our requirements on convergence demand this to happen within finite
time, which is very important for numerical/experimental
investigations. In these cases, contrary to a rigorous mathematical
framework and due to physical limitations one cannot ask for time going
to infinity or infinitely small length intervals. By making such
\textit{finite-size} assumptions on the dynamics one may include
among the detected invariant sets homoclinic tangencies and Newhouse
attractors which support some invariant measure at least within finite
scales, thus being indistinguishable under finite resolution from more
general ``real'' attractors~\cite{GST96, RMG09}.

We will focus on the case where there is only a finite number of
coexisting attractors. This is not a restriction, because for
compact spaces the finiteness of the number of effective attractors
follows. Indeed, it is only possible to fit a finite number of non
overlapping balls of radii bounded from below in a compact
space. Furthermore, for the case of randomly perturbed dynamics we
shall deal with it can be proven that the system has only a finite
number of invariant \textit{physical}
measures~\cite{Ara00}. Therefore, we represent the set of coexisting
effective attractors by $\{\Lambda_{i}\}_{i=1}^{N}$, a family of
\textit{pairwise disjoint} compact sets, i.e. $\Lambda_{i} \cap
\Lambda_{j} = \emptyset$, for $i \neq j$. Another important fact is
that we also assume that the union of the basins of attraction covers
every point of the whole phase space, up to a zero Lebesgue measure
set. So we write
\begin{equation}
\label{eq.basins}
m\left ( M \backslash \bigcup_{i=1}^{N}W^{s}(\Lambda_{i})\right) = 0,
\end{equation}
where $m$ denotes Lebesgue measure and $W^{s}(\Lambda_{i})$ the
basin of attraction of $\Lambda_{i}$. This plays a very important role
in the definition of the hopping process between different attractors,
because the trajectories are always expected to converge to some
attractor. The boundary between basins of attraction is a zero
Lebesgue measure component, the so-called \textit{basin boundary},
which we denote by $\partial$. The basin boundary plays a fundamental
role in the hopping process, as we shall see in what follows.

\label{sec.randpert}

\section{Random perturbations}

We now perturb the dynamics exhibiting multiple attractors by assuming
\textit{physical random perturbation}; see~\cite{Ara00} and Appendix D
of~\cite{BDV05} for a formal definition. Roughly speaking we add
bounded random uniformly distributed noise to the dynamics. That is,
given the deterministic system $f$ defined as before, we consider the
dynamical system
\begin{equation}
\label{eq.perturbed}
F(x_{j}) = f(x_{j}) + \varepsilon_{j},
\end{equation}
with $||\varepsilon_{j}|| < \xi$, where $\varepsilon_{j}$ is the
random vector of noise added to the deterministic dynamics at the
iteration $j$, and $\xi$ is its maximum amplitude. We require the
noise to asymptotically cover uniformly a ball around the unperturbed
dynamics, representing the idea that the perturbation has no
preferential direction and amplitude. The orbit thus jumps from $x$ to
$f(x)$ but misses the point at random with the conditional probability
of finding the perturbed orbit in an $\xi$-neighbourhood of $f(x)$
given $x$, see~\cite{JKR12} for a comprehensive treatment of this
topic.

\section{Escape}


If the amplitude of the perturbations is small enough, an orbit in the
domain of attraction approaches the attracting set, wanders around
without escaping and is expected to be trapped there forever. Although
the trajectory may seem very intricate, it is actually well described
from a statistical perspective. In these cases, one has a unique
invariant ergodic probability distribution representing a given
attracting set~\cite{Ara00}. If the system is \textit{stochastically}
stable, such distributions for the randomly perturbed system approach
those of the deterministic one as the amplitude of the perturbations
decreases to zero.  The dynamics inside the basin can be described as
that of a closed system if the amplitude of the perturbations is small
enough~\cite{RGM10}. When the amplitude of the noise increases beyond
a threshold $\xi_{0}$ the attracting sets lose their stability. This
effect can be seen as the introduction of a hole $I_{\partial} =
I_{\partial}(\xi)$ in the basin by which the orbits can escape from
the domain of attraction; see \cite{RGM10} and further
references therein for the general setting. Under some assumptions it
is possible to estimate the size of such a hole, or its measure
$\mu(I_{\partial}) > 0$ \cite{RGM10}.  For one dimensional systems
rigourous results in this direction have been obtained with a
different approach~\cite{ZmH07}.

\section{\label{sec.hop}Hopping process}

Now we are ready to translate the problem of noise induced escape from
\textit{pseudo attractors} into that of a closed system with a hole
$I_{\partial}$, or a recurrence problem. We call \textit{pseudo
attractors} the sets where the orbits remain trapped for some amount
of time before escaping due to noise. Rigourously speaking they are
not attractors or attracting sets, since the invariance condition is
not fulfilled. In our context, a pseudo attractor $A$ is a
\textit{quasi-invariant set} when the amplitude of the random
perturbations is increased beyond $\xi_{0}$.  With the assumption
above we can describe our dynamics and the escape from a single
attractor as $x_{j+1} = F(x_{j})$ if $x_{j} \in A$ or \textit{escape}
if $x_{j} \in I_{\partial}$. We do not define the dynamics in
$I_{\partial}$ as it is irrelevant to our discussion, hence when the
orbit falls into $I_{\partial}$ we stop considering it. However, we
allow the trajectory to come back from the hole to $A$. If so, we
restart the process of counting the time in $A$ by neglecting the
number of iterations that it had spent in $I_{\partial}$.

Similar arguments apply to systems with many coexisting pseudo
attractors $A_{i}$ for which Eq.~(\ref{eq.basins}) holds. In such
dynamics, when a trajectory falls into the $i$th hole
$I_{\partial_{i}}$ there is the possibility of swapping basins. Using
a Markov assumption we argue this to be equivalent to restarting the
process. Although for the $i$th hole there is a distinct measure
$\mu_{i}(I_{\partial_{i}})>0$, according to our assumption we treat
all holes qualitatively in the same way.  Ignoring the dependence on
$i$ we simplify the recurrence in probability space to the $i$th
interval by dropping the index $i$. We are thus characterising the
dynamics in terms of a representative hole $I_{\partial}$ with average
measure $\mu(I_{\partial})$. Correspondingly we reduce the sojourn
time distribution of the hopping process to the statistics of the time
intervals that a random orbit takes to access the representative hole
$I_{\partial}$. Furthermore, we assume the general \textit{basin
  property} to hold, which tells us that up to a set of zero Lebesgue
measure the time averages of orbits in the basins of attraction
converge to the space average with respect to the invariant measures
supported on the attractors; see Chap 1.6 in~\cite{BDV05}.

\section{\label{sec.stick}Pseudo stickiness}

Let us now look further at the microscopic dynamics in order to
understand the overall statistical behaviour of the noise induced
hopping process between different attractors. In particular we shall
explore its analogy with non-hyperbolic Hamiltonian dynamics where
\textit{stickiness} plays a fundamental role for explaining the
statistical dynamics.

To set the scene let us forget about the noise for the moment. Recall
that Hamiltonian non-hyperbolic dynamics is characterised by elliptic
orbits, whose eigenvalues are purely imaginary. These orbits are
surrounded by complex structures formed by marginally stable periodic
orbits, known as Kolmogorov-Arnold-Moser (KAM) invariant tori or
islands, as well as regions of chaotic motion. Large islands are
surrounded by smaller ones which, on the other hand, are surrounded by
even smaller ones, repeating this pattern on smaller scales \textit{ad
  infinitum}. Trajectories starting in the chaotic region exhibit
\textit{intermittent dynamics}: they spend long sporadic periods of
time performing almost regular motion near the borders of the islands
before escaping to the chaotic sea again. Even small islands can have
a great impact on the dynamics of an orbit. Given the hierarchical
structure of the phase space, when an orbit eventually escapes from
the neighbourhood of an island it may spend some time wandering in the
chaotic sea before it gets \textit{trapped} once more by the same or
another island. This effect, generally known as
\textit{stickiness}~\cite{Ott02}, slows down the dynamics. Among its
statistical signatures one typically observes power-law decay of
correlations and anomalous diffusion~\cite{KRS08}.

\textit{Uniformly hyperbolic} dynamics, on the other hand, is
characterised by exponential-like laws. Roughly speaking a system is
called hyperbolic if at each point on the attracting set distances are
contracted or expanded with exponential rate. If the rate of
convergence does not depend on the point, the system is called
uniformly hyperbolic~\cite{BDV05}. In what follows we argue that, from
a statistical point of view, in our case \textit{the presence of
random perturbations destroys uniformly hyperbolic behaviour.} That is,
the perturbations destroy uniform contraction and expansion rates,
therefore exponential statistical signatures are lost. Furthermore,
when the noise amplitude is set above a threshold, the orbits can
escape from the attracting sets as explained in the previous
section~\textbf{Escape}. The general statistical effect
is similar to that observed in non-hyperbolic Hamiltonian
systems. Namely, the pseudo attractors behave in a manner similar to
the KAM islands, where the orbits perform an almost regular motion for
a limited time interval.  The presence of noise furthermore washes out
fine scale structures of the phase space. Thus, the trapping regions
of small attractors have less but non-negligible importance, since the
orbits might stay inside them only for a short time by performing
almost regular motion before escaping again. Once an orbit escapes
from a pseudo attractor, it undergoes an erratic motion until it falls
again into the same or another trapping region. Although some of the
trapping regions may be very small, yet they have great influence on
the statistical characterisation of the dynamics because, just like
small KAM islands in the case of non-hyperbolic dynamics, every
pseudo-attractor has a stickiness-like effect. An important difference
nevertheless is that for the dissipative case, the attractiveness to a
nearly invariant sets determines the type of diffusion. The mean
square displacement is thus expected to show a slower diffusive
dynamics compared to Hamiltonian systems. 

\section{\label{sec.s-hyp} Sojourn time distribution and
  Hyperbolicity}

In the previous sections we focused on the connection between a
hopping process and escape in a dynamical system with holes. As a
consequence, the sojourn time distribution for the hopping process
given by the distribution of escape times $P(t)$ for a system with
holes depends on the dynamics in the pseudo attractors (\textit{i.e.}
the sets $A_{i}$) governed by their hyperbolic properties. We consider
two ``extreme types'' of dynamics: on the one side, the escape of
orbits from sets in uniformly hyperbolic dynamical systems has been
shown to follow an exponential time distribution. On the other side,
escape in Hamiltonian systems with mixed phase space yields power-law
tails~\cite{AlT08, ASC04, Ott02, BaB90}.

Now suppose that in a given dynamical system we could somehow control
``how hyperbolic'' it is.  We might then switch the escape time
distribution between $P(t) \approx a e^{-\alpha t}$ and $P(t) \approx
b t^{-\beta}$, where the parameters $a$ and $b$ depend on the
hyperbolicity of the dynamics. They are determined by the dynamics in
the pseudo attractors, or more generally, in the set with a hole from
where the trajectories escape.  For uniformly hyperbolic systems the
parameter $a$ is large and the dynamics in the pseudo attractor has
hyperbolic characteristics. Therefore, we have a hyperbolic recurrence
time distribution to $I_{\partial}$, and the asymptotic decay of the
corresponding escape times is exponential. On the other hand, when the
non-hyperbolic component of the dynamics is increased, the parameter
$b$ gains importance and the diffusion of the random orbit in the
support of the conditionally invariant measure\footnote{The
  conditional measure is defined such that, for each iteration,
  when the set $A$ loses a fraction of its orbits to the hole, we
  renormalise its measure by what remains in $A$; see\cite{RGM10} for
  details.} experiences a stickiness effect, resulting in a slower
distribution of recurrence times to $I_{\partial}$ with a power-law
tail. Such an increase of non-hyperbolic characteristics under
parameter change may be the result of homoclinic tangencies with
highly non-uniformly hyperbolic properties~\cite{BDV05, GST96}. Since
we deal with dynamics under finite resolution, we cannot distinguish
them from the other attractors. Note that this behaviour should be
independent of the noise amplitude within some range of it, because
its amplitude will control the number of pseudo attractors, but the
type of escape should be controlled by the hyperbolicity of the
system. In the next section we present numerical evidence supporting
our arguments, showing that for systems close to the non-hyperbolic
regime the escape time distribution indeed has the power law signature
of non-hyperbolicity rather than being exponential as expected for
uniformly hyperbolic dynamics.

\section{\label{sec.num} Numerical results}

We illustrate our results by simulations of the perturbed system
defined by $F(x_{j}, y_{j}) = f(x_{j}, y_{j}) + (\varepsilon_{x,j},
\varepsilon_{y,j})$ with uniformly distributed \textit{i.i.d} random
noise. For $f$ we choose the single rotor map~\cite{Zas78}
\begin{equation}
\label{eq.single-rotor-map} 
f \left(\begin{array}{c} x_{j} \\ y_{j}
\end{array}\right) = \left( \begin{array}{c} x_{j} + y_{j} \: \mod 2\pi\\ 
(1-\nu)y_{j} + f_{0} \sin(x_{j} + y_{j}) \end{array} \right),
\end{equation} 
with $x \in [0,2\pi]$, $y \in \mathbb{R}$ and damping parameter $\nu
\in [0,1]$. When $\nu \neq 0$ the dynamics is dissipative. In the
strongly dissipative limit $\nu \rightarrow 1$ this model shows
uniformly hyperbolic statistical properties, at least from
the perspective of effective
attractors~\cite{FGH96}. Conversely, when $\nu \rightarrow 0$ the
dynamics approaches the non-hyperbolic Hamiltonian limit, and under
finite resolution dynamics there is an increase of the number of
periodic attractors~\cite{FGH96, RMG09}. For $\nu = 0$ we recover the
area preserving standard map with Hamiltonian
dynamics~\cite{Chi79}. Therefore, we can think of $\nu$ as a control
parameter measuring how far the dynamics is away from the
non-hyperbolic regime. We use $f_{0} = 4.0$, which results in multiple
attractors when $\nu \neq 0$~\cite{FGH96}. At this parameter value and
$\nu =0$ the standard map displays superdiffusion, due to the
existence of accelerator modes \cite{MaR14}.

If we evolve our system under the presence of random noise beyond a
certain amplitude $\xi \geq \xi_{0}$ the attracting sets lose their
stability, as discussed in the
section~\textbf{Escape}. Note that each attractor may have
a different value of minimum noise amplitude such that escape takes
place, which is proportional to the size of their basins of
attraction. We choose as a global $\xi_{0}$ the minimum value for the
escape from the largest trapping region. For $\xi \geq \xi_{0}$ escape
from the attracting sets consequently gives rise to diffusion of
trajectories through the phase space.
\begin{figure}[t]
    \subfigure[]{\label{fig.pdf}\includegraphics[width=.85\columnwidth]{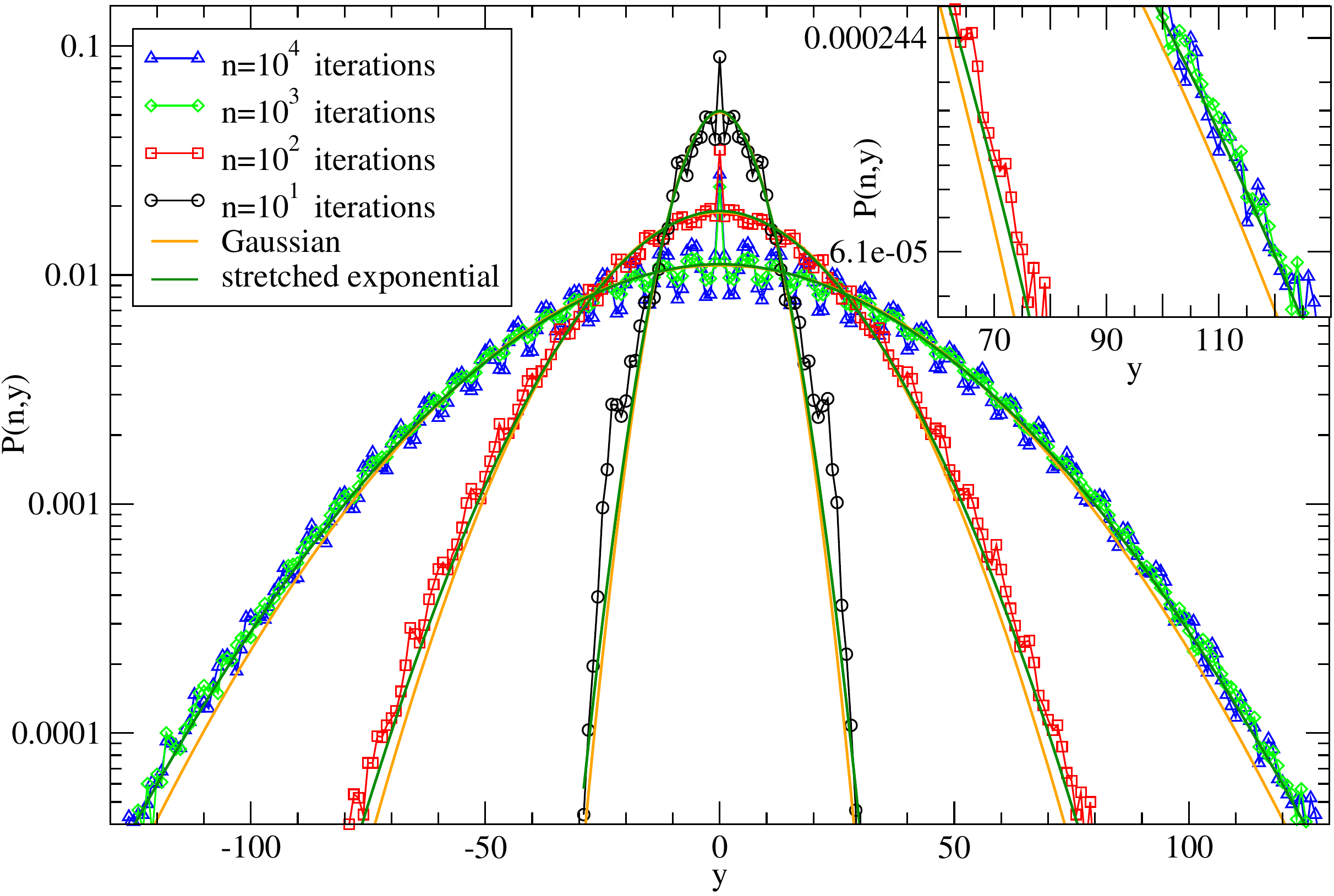}}
    \subfigure[]{\label{fig.criteria}\includegraphics[width=.85\columnwidth]{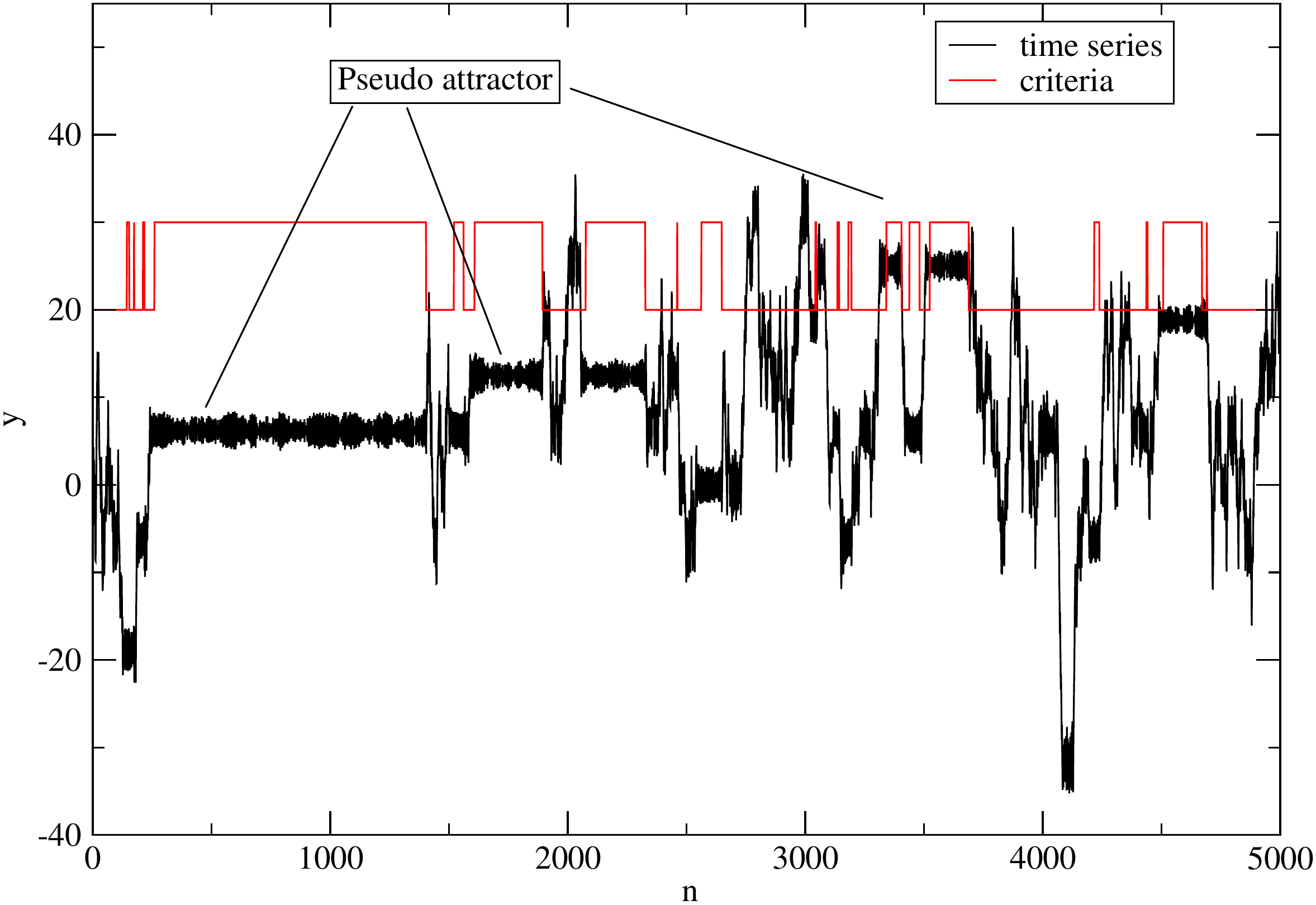}}
  
\vspace*{-0.3cm}
  \caption{(a) Probability density function $P(n,y)$ at
    position $y$ for different iteration numbers $n$. An ensemble of $10^{6}$ random initial
    conditions uniformly distributed around
    $x=y=0$ was iterated by the map Eq.~(\ref{eq.single-rotor-map})
    randomly perturbed by noise of level $\xi
    = 0.06$ and dissipation $\nu = 0.002$.  The lower (orange) lines
    display fits with Gaussian distributions for the three smaller
    $n$, the
    upper (dark green) lines are stretched exponential fits with
    Eq.~(\ref{eq:strexp}). The inset shows a blowup of two
    tails. (b) The black graph depicts a representative
    time series of the noisy system for $\nu = 0.02$ and $\xi = 0.2$.
    The corresponding result by our eigenvalue
    criterion to identify pseudo attractors (see text) is given by the
    red line.  The plateaus at $y=30$ reveal pseudo attractors, which
    coincide with the visual identification of localisation in the
    time series.}
  \label{fig.pdf-criteria}
\end{figure}

Fig.~\ref{fig.pdf} shows the time dependence of the $y$-position
probability density function of such a process. It confirms our
hypothesis that diffusion of trajectories induced by random
perturbations indeed takes place. While at first view the included
fits to Gaussian distributions seem to match well to the simulation
data, the inset shows deviations in the tails especially for long
times. This deviation will be explained later on by matching
the data with a stochastic theory. Note also the existence of a
periodic fine structure, which reflects the spatial distribution of
the attracting sets along the $y$ axis \cite{FGH96}. Analogous results
have been obtained for simulations under different levels of random
noise, for different dissipation parameters $\nu$, and also for
different values of $f_0$.

For general systems a rigourous investigation of the sojourn time
distribution and the identification of pseudo attractors can be a very
difficult task~\cite{Ara00, ZmH07}. Even from the numerical point of
view the fact that, \textit{a priori}, neither the physical nor the
conditionally invariant measures are known can represent an obstacle
to the identification of pseudo attractors.  A way to detect whether
an orbit is trapped in the trapping region of some pseudo attractor
for a period of time is given in terms of finite-time Lyapunov
exponents. Equivalently, one can calculate the eigenvalues of the
Jacobian matrix of $F$ along the orbit. As a consequence of
meta-stability of the pseudo attractors, while an orbit remains
trapped the maximum eigenvalue of the Jacobian has, on average,
magnitude less than one; see Theorem V1.1 in~\cite{Kif86} for a
rigourous discussion on characteristic exponents in the case of random
transformations.  Fig.~\ref{fig.criteria} illustrates our criterion
for the random dynamical system Eq.~(\ref{eq.single-rotor-map}) where
we have, without loss of generality, plotted $y=30$ when a pseudo
attractor is identified and $y=20$ otherwise. Also without
loss of generality we only consider trajectories that remain trapped
for more than $20$ iterations.

\begin{figure}[t]
  \begin{center}
    \subfigure[]{\label{fig.histnoise}\includegraphics[width=.85\columnwidth]{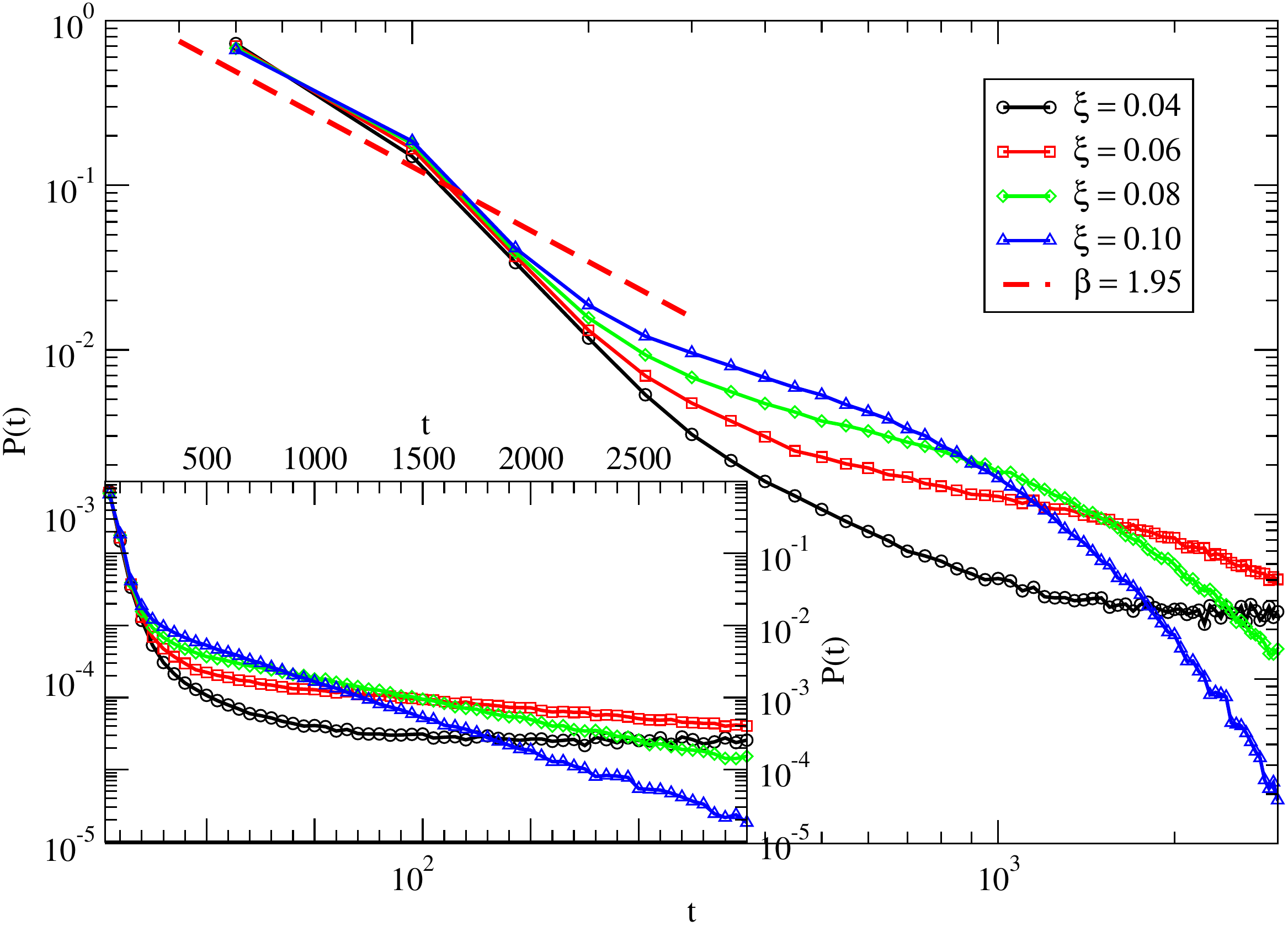}}
    \subfigure[]{\label{fig.moment}\includegraphics[width=.85\columnwidth]{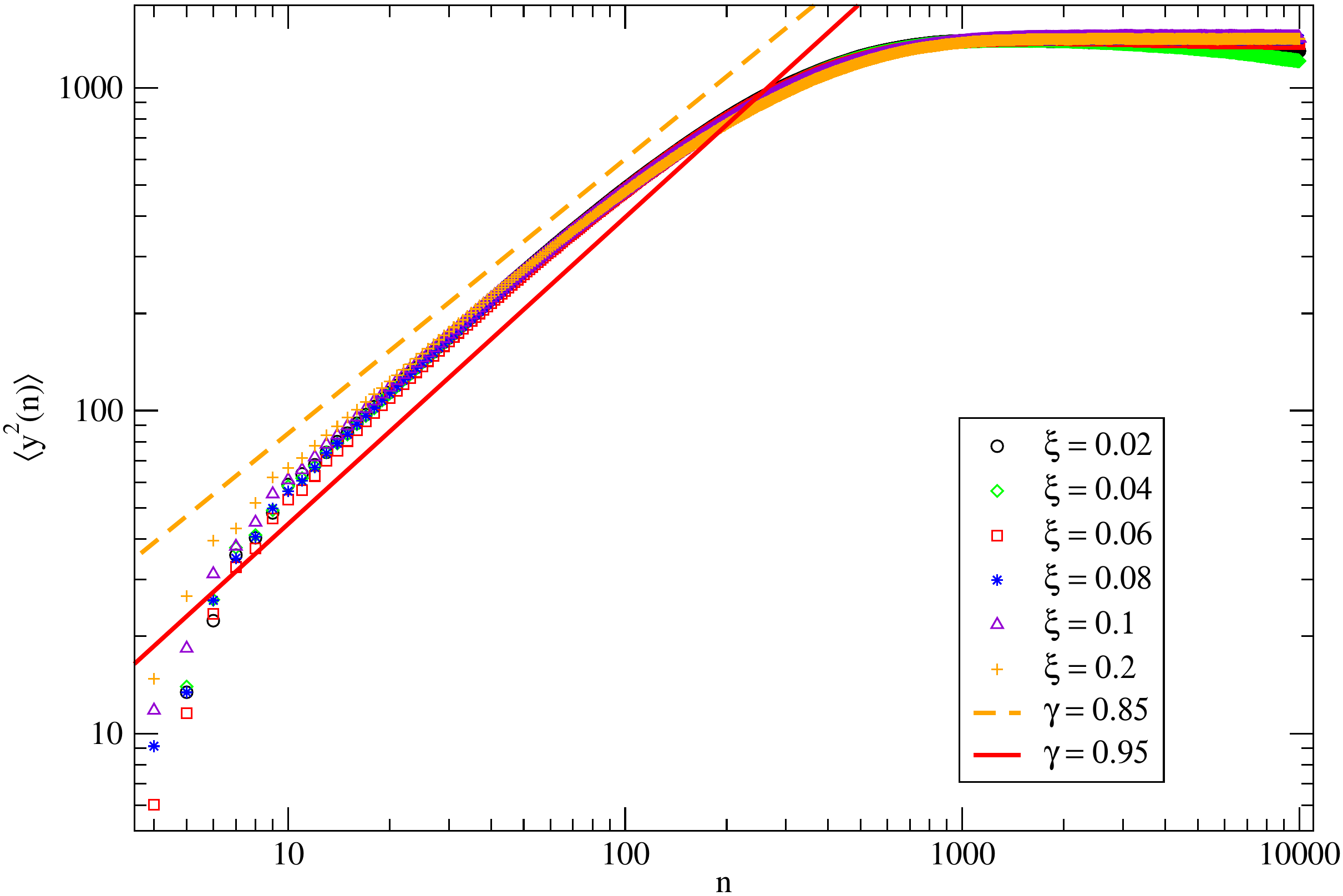}}
  \end{center}
%
\caption{(a) Double-logarithmic plot of the
probability distribution $P(t)$ of escape times $t$
  for an orbit to stay trapped in a pseudo attractor for $n < t$. The
  map~Eq.~(\ref{eq.single-rotor-map}) was iterated $10^{9}$ times for
  dissipation $\nu = 0.002$ and different values of the noise
  amplitude
  $\xi$. The dashed line represents a power law
    decay with exponent $\beta=1.95$. The inset shows the
    corresponding semi-logarithmic plot. (b) Mean square
  displacement $\langle y^2(n)\rangle$ for the coordinate $y$ as a
  function of time $n$.
  An ensemble of $10^{6}$ initial conditions was iterated by the
  map Eq.~(\ref{eq.single-rotor-map}) for different amplitudes $\xi$ of
  random noise and fixed small dissipation $\nu =
    0.002$. The lower bold line corresponds to a power law with
    exponent $\gamma=0.95$, the upper dashed line to an exponent
    $\gamma = 0.85$.}
\end{figure}

Once a proper identification of the different dynamical regimes,
\textit{i.e.}  trapped or wandering, is obtained, we are ready to
statistically analyse these different behaviours. We start by
computing the probability distributions for the times an orbit stays
trapped for $n < t$ iterations in a pseudo attractor. For a range of
larger values of $\nu$ in our simulations we observe a predominantly
exponential escape, as was to be expected~\cite{Ott02, BaB90, AlT08,
  ASC04}. However, when the damping is decreased below $\nu =
  0.02$ the probability distribution is roughly described by a power
  law, similar to the case of non-hyperbolic Hamiltonian
  dynamics~\cite{AlT08}. In Fig.~\ref{fig.histnoise} we show the
  probability distributions of escape times from pseudo attractors, or
  equivalently, the first recurrence time distributions to
  $I_{\partial}$,
  for fixed small dissipation $\nu$ but different noise amplitudes
  $\xi$. Approximately up to times $t<300$ the escape time
  distributions match reasonably well to power laws with exponents
  around $\beta=1.95$ as shown in the figure. This will be
  justified later by matching all data consistently with a theoretical
  prediction. The value is in agreement with the range of
  exponents $1.5\le \beta \le 3$ obtained
  analytically for trapping regimes in bounded Hamiltonian
  systems~\cite{Ven09}.

Although the precise value of the noise escape threshold $\xi_{0}$
depends on the parameters $f_{0}$ and $\nu$, for amplitudes $\xi \geq
\xi_{0}$ the existence of a power law decay is independent from the
amplitude of the noise. This is not shown here but observed in further
simulations.  When we decrease $\xi$ the orbit typically takes longer
to escape, consequently the probability distributions are stretched to
longer times. In Fig.~\ref{fig.histnoise} we observe a cross-over to
exponential laws which changes with $\xi$, as is highlighted
by the inset. The most important result of this analysis is that when
the dynamics is near the non-hyperbolic Hamiltonian limit,
\textit{i.e.} for small dissipation parameters $\nu$, the behaviour of
diffusive trajectories indeed has, from the statistical point of view,
non-hyperbolic characteristics. This is what we shall address next.

In order to understand the type of diffusion process we are dealing
with, we computed the mean square displacement $\langle y^2(n)\rangle$
for the coordinate $y$, the relevant one for diffusion, as a function
of time $n$.  The two lines shown in Fig.~\ref{fig.moment}
represent power laws $\langle y^{2}(n)\rangle\sim n^{\gamma}$ with
exponents $\gamma < 1$. They reveal power law behaviour for the data
up to approximately $t<300$ by providing upper and lower bounds for
the exponents. For the corresponding subdiffusive hopping process
among the different basins the power laws persist independently of
$\xi$ but with a slightly varying exponent. Changing other parameters
such as $\nu$ typically generates the same behaviour. This finding is
in agreement with our analogy to non-hyperbolic Hamiltonian dynamics
generating stickiness to pseudo attractors as discussed in
Section~\textbf{Pseudo Stickiness}. Note that for $t>1000$ all power
law exponents of $\langle y^{2}(n)\rangle$ are close to zero. This is
due to the fact that the fastest particles have reached the region in
phase space where the pseudo attractors of the map cease to exist
\cite{FGH96} meaning they cannot move any further, and trivial
localization sets in.

In the area preserving standard map superdiffusion
  has successfully been modeled by stochastic \textit{Continuous Time
    Random Walk (CTRW) theory} \cite{ZuKl94}. As our randomly
  perturbed dissipative model displays subdiffusion, here we test the
  subdiffusive CTRW version put forward in
  Refs.~\cite{GeTo84,MKl00,Kor07} to explain our
  simulation results. This theory predicts that if the mean square
  displacement exhibits a power law with exponent $\langle
  y^{2}(n)\rangle\sim n^{\gamma}$, the respective escape (or waiting)
  time distribution must be $P(t)\sim t^{-(\gamma+1)}$ on
  corresponding time scales. It furthermore predicts that the position
  distribution function of the subdiffusive process must approximately
  be of the stretched exponential form.\footnote{In detail the asymptotic CTRW results for
  $P(n,y)$ look a bit different, cf.\ Eq.~(51) in Ref.~\cite{Kor07}. But
  we have checked that the stretched exponential dominates the
  expression for our $\gamma$ and at least large $y$.} 
\begin{equation}
P(n,y)\sim\exp\left(-c(n) y^{2/(2-\gamma)}\right) \label{eq:strexp}\:.
\end{equation}
The lower straight line in Fig.~\ref{fig.moment} representing a power
law with exponent $\gamma=0.95$ matches well to the mean square
displacement of $\xi=0.06$. The dashed line in
Fig.~\ref{fig.histnoise} yields the corresponding power law with
exponent $\gamma+1=1.95$ as predicted by CTRW theory, which matches
well to the numerical result for the escape time distribution for the
same $\xi=0.06$ in the regime of $t<300$ where the system is
subdiffusive. Finally, the stretched exponential fits for $\xi=0.06$
in Fig.~\ref{fig.pdf} have all been performed with
Eq.~(\ref{eq:strexp}) by using the very same value of $\gamma$.
Evidently, these fits match much better to the numerical results in
the tails than the corresponding Gaussian distributions, at least for
long enough times. We thus conclude that the subdiffusive CTRW of
Refs.~\cite{GeTo84,MKl00,Kor07} consistently explains our numerical
findings, thus confirming theoretically that our randomly perturbed
dissipative dynamics generates a subdiffusive process that is
well-known in stochastic theory. This is quite surprising, as we did
not take the strongly non-uniform distribution of pseudo attractors
along the $y$ axis into account but just averaged over all of them by
performing a kind of mean field approximation.

\section{Conclusion}
\label{sec.concl}

We have investigated the hopping process of points generated
by randomly perturbed dissipative dynamics. We have set up a
theoretical framework that describes escape in terms of a closed
system with a hole.  Escape occurs when the support of the conditional
invariant measure of one pseudo attractor overlaps with the
neighbourhood of another basin boundary. In this setting the sojourn
time distribution becomes the recurrence time distribution of the
orbit wandering to a hole. We then showed by simulations that for the
randomly perturbed weakly dissipative single rotor map the
distribution of sojourn times is described by a power law up to
relevant time scales, in contrast to an exponential distribution for
strong dissipation. We found that the hopping process among different
basins is subdiffusive for a wide range of perturbation
strengths. Using only the subdiffusive power law exponent as a fit
parameter, we showed that stochastic CTRW theory consistently explains
all of our simulation data by revealing stretched exponential tails in
the position distribution function. We conclude that bounded random
perturbations generate a kind of non-hyperbolic stickiness in the
diffusion process for the considered dissipative dynamics which leads
to non-Gaussian position distributions, power laws in the escape time
distributions, and subdiffusion. It would be interesting to
investigate whether similar phenomena occur in other diffusive
randomly perturbed deterministic dynamical systems.

\acknowledgements The research leading to these results has received
funding from the European Research Council under the European Union's
Seventh Framework Programme (FP7/2007-2013) / ERC grant agreement
n$^\circ$~267087. RK thanks the MPIPKS Dresden for hospitality.


\end{document}